\documentclass[twocolumn,trackchanges]{aastex62}
\usepackage{graphicx}
\turnoffedit
\newcommand\pcc{\;{\rm cm}^{-3}}
\newcommand\Msun{\; M_{\odot}}
\newcommand\kms{\; {\rm km}\;{\rm s}^{-1}}

\newcommand\cm{\;{\rm cm}}
\newcommand\yr{\; {\rm yr}}

\newcommand\Myr{\;{\rm Myr}}

\newcommand\pc{\;{\rm pc}}
\newcommand\kpc{\;{\rm kpc}}

\newcommand\sfrunit{\Msun \kpc^{-2} \yr^{-1}}

\newcommand\Surf{\Msun\;{\rm pc^{-2}}}

\newcommand\Kel{\;{\rm K}}

\newcommand\simgt{\lower.5ex\hbox{$\; \buildrel > \over \sim \;$}}
\newcommand\simlt{\lower.5ex\hbox{$\; \buildrel < \over \sim \;$}}

\newcommand\rbrackets[1]{\left({#1}\right)}

\newcommand\abrackets[1]{\left\langle{#1}\right\rangle}

\newcommand\vel{\mathbf{v}}
\newcommand\Bvec{\mathbf{B}}

\newcommand\xhat{\hat{\mathbf{x}} }
\newcommand\yhat{\hat{\mathbf{y}} }
\newcommand\zhat{\hat{\mathbf{z}} }

\newcommand\Xhat{\hat{\mathbf{X}}}
\newcommand\Yhat{\hat{\mathbf{Y}}}
\newcommand\Zhat{\hat{\mathbf{Z}}}

\newcommand\rTE{r_{TE}}

\newcommand\vturb{v_{\rm turb}}

\newcommand\smax{s_{\rm max}}
\newcommand\smin{s_{\rm min}}
\newcommand\lmax{\ell_{\rm max}}
\newcommand\lmin{\ell_{\rm min}}
\newcommand\fsky{f_{\rm sky}}
\newcommand\REB{\mathcal{R}_{ EB}}
\newcommand\planck{\emph{Planck}}
\newcommand\wmap{\emph{WMAP}}
\newcommand\SigmaSFR{\Sigma_{\rm SFR}}
\shorttitle{Dust Polarization Maps from TIGRESS}
\shortauthors{Kim, Choi, \& Flauger}
\received{2019 January 20}
\revised{2019 June 12}
\accepted{2019 June 13}
\submitjournal{Astrophysical Journal}

\begin{document}

\title{Dust Polarization Maps from TIGRESS: $E/B$ power asymmetry and $TE$  correlation}

\author[0000-0003-2896-3725]{Chang-Goo Kim}
\affiliation{Department of Astrophysical Sciences, Princeton University, Princeton, NJ 08544, USA}
\affiliation{Center for Computational Astrophysics, Flatiron Institute, New York, NY 10010, USA}
\author{Steve K. Choi}
\affiliation{Department of Astronomy, Cornell University, Ithaca, NY 14853,
USA}
\affiliation{Joseph Henry Laboratories of Physics, Jadwin Hall, Princeton
University, Princeton, NJ 08544, USA}

\author[0000-0001-9527-8678]{Raphael Flauger}
\affiliation{University of California, San Diego, La Jolla, CA 92093, USA}

\email{cgkim@astro.princeton.edu}

\begin{abstract}
We present the first large set of all-sky synthetic dust polarization maps derived directly from a self-consistent magnetohydrodynamics simulation using the TIGRESS framework. Turbulence in this simulation is predominantly driven by supernova explosions, with rates that are self-consistently regulated by feedback loops. The simulation covers both the outer scale and inertial range of turbulence with uniformly high resolution. The shearing-box utilized in the simulation in concert with resolved supernova-driven turbulence enables to capture generation, growth, and saturation of both turbulent and mean magnetic fields. We construct polarization maps at 353 GHz as seen by observers inside a model of the multiphase, turbulent, magnetized interstellar medium (ISM). To fully sample the simulated ISM state, we use 350 snapshots spanning over $\sim350\Myr$ (more than six feedback loops) and nine representative observers. The synthetic skies show a prevalent $E/B$ power asymmetry ($EE>BB$) and positive $TE$  correlation in broad agreement with observations by the \planck{} satellite. However, the ranges of $EE/BB\sim1.4-1.7$ and $TE/(TT\cdot EE)^{1/2}\sim0.2-0.3$ are generally lower than those measured by \planck{}. We find large fluctuations of $E/B$ asymmetry and $TE$ correlation depending on the observer's position and temporal fluctuations of ISM properties due to bursts of star formation. The synthetic maps are made publicly available to provide novel models of the microwave sky. 
\end{abstract}

\vskip 2cm
\section{Introduction}\label{sec:intro}

Interstellar magnetic fields are believed to play a crucial role in controlling star formation either by directly providing additional support to the self-gravitating entities on large and small scales \citep{1987ARA&A..25...23S} or by altering the characteristics of turbulence \citep{2004ARA&A..42..211E,2007ARA&A..45..565M}. As a consequence, understanding interstellar magnetic fields is of paramount importance in the field of star formation and the study of the interstellar medium (ISM).

Measurements of the polarization of the microwave sky by the \planck{} satellite have recently delivered a wealth of information about interstellar magnetic fields. For example, because non-spherical interstellar dust grains are expected to align tightly with local magnetic fields \citep[e.g.][]{2008MNRAS.388..117H,2015ARA&A..53..501A}, the polarized thermal dust emission directly traces the interstellar magnetic fields shaped by a variety of nonlinear physical processes, including large scale energy injection from star formation feedback, turbulence cascade, and galactic dynamo \citep[][for a review]{2015ASSL..407..483H}.
 
\citet[][PXXX hereafter]{2016A&A...586A.133P} characterized the $E$-mode and $B$-mode angular power spectra of the 353~GHz polarization maps over the multipole range $40<\ell<600$. $E$ and $B$ modes form a basis of linear polarization, and are expected to be equal for randomly oriented polarization \citep[e.g.,][]{1997PhRvD..55.7368K,1997PhRvD..55.1830Z,2001PhRvD..64j3001Z}. However, the observations show that the $E$-mode (gradient) power is about a factor two larger than the $B$-mode (curl) power \citep[see also][ for recent updates]{2018arXiv180104945P}.  In the same analysis, $E$-mode polarization shows a positive correlation with temperature anisotropy ($TE$ correlation). $EB$ correlations are statistically insignificant but there are recent claims of a weak positive $TB$ correlation \citep{2018arXiv180104945P, 2019arXiv190510471B}. 

These \planck{} observations provide new constraints that models of the local ISM must satisfy. An asymmetry in the $E$- and $B$-mode powers may emerge naturally from the perspective of magnetohydrodynamic (MHD) turbulence in the ISM because turbulence is not random motion, but follows a specific scaling relation \citep[e.g.,][]{1941DoSSR..30..301K,1972LNP....12...41B,2011PhRvL.106g5001B} with scale-dependent anisotropy in the presence of strong magnetic fields \citep[e.g.,][]{1995ApJ...438..763G,2000ApJ...539..273C,2003MNRAS.345..325C,2006PhRvL..96k5002B}.

Using an analytic description of the {\it statistical} properties of MHD turbulence, \citet{2017ApJ...839...91C} claimed that the observed $E/B$ power asymmetry and positive $TE$  correlation are not easily reproduced by simple MHD turbulence prescriptions decomposed into linear MHD waves. The model adopted an arbitrary functional form for an anisotropy, and the authors questioned a narrow parameter range for an anisotropy parameter (as well as the plasma beta defined by the ratio between thermal and magnetic pressures) required by the observed constraints. \citet{2017MNRAS.472L..10K} performed a similar analysis and reached a different conclusion. \citet{2017MNRAS.472L..10K} replaced the anisotropy parameter with Alfv\'en Mach number using an anisotropy function derived from the Goldreich-Sridhar theory \citep[][]{1995ApJ...438..763G,2003MNRAS.345..325C,2012ApJ...747....5L}. In this model, the observed $E/B$ power asymmetry can be reconciled with sub-Alfv\'enic turbulence (magnetic energy is larger than turbulence kinetic energy), while the positive $TE$ correlation additionally requires uncorrelated magnetic field and density \citep{2018MNRAS.478..530K}. The desired parameter range for turbulence is still narrow (especially if as suggested in \citet{2017MNRAS.472L..10K} slow and Alfv\'en waves dominate), but these conditions appear plausible for the high-latitude, low-density ISM analyzed in PXXX. \citet{2018MNRAS.478..530K} further suggested that an asymmetry would exist in synchrotron polarization as well in sub-Alfv\'enic turbulence, consistent with both the \wmap{} \citep{2007ApJS..170..335P} and \planck{} data. 

These studies give some hints how MHD turbulence could be responsible for the observed results. However, a direct application of this analysis to the real ISM is not entirely straightforward in part because it is difficult to develop an analytic theory of MHD turbulence for the multiphase ISM. The real ISM is subject to a variety of cooling and heating processes, resulting in distinct thermal phases with typical densities and temperatures that span many orders of magnitude \citep[e.g.,][]{1977ApJ...218..148M,2003ApJ...587..278W,2017ApJ...846..133K}. 

A complementary approach to understand the observed dust polarization is to use direct numerical simulations of MHD turbulence for the multiphase ISM. \citet{2018PhRvL.121b1104K} investigated MHD simulations of driven multiphase turbulence in a periodic box. In these simulations, turbulence is driven by large-scale random solenoidal forcing, and the simulations include major cooling and heating processes for the cold ($T\sim 10^2 \Kel$) and warm ($T\sim 10^4 \Kel$) medium. The synthetic dust maps based on these simulations of the two-phase ISM show good agreement with the observed constraints. The results depend on the ``turbulence parameters'' (e.g., Alfv\'en Mach number, sonic Mach number), which are given as initial conditions in this type of simulations. For strong magnetic fields, the simulations show a weak dependence on the choice of projection with respect to the mean field direction. More targeted simulations informed by observations and/or large-scale simulations are required to refine the comparison with data.

Polarized microwave emission also plays a crucial role for next-generation cosmic microwave background (CMB) experiments that aim to detect the unique B-mode signal induced by primordial gravitational waves because the primordial B-mode signal may be weaker by many orders of magnitude than the galactic foreground signals from dust and synchrotron \citep{2015JCAP...12..020C, 2016ARA&A..54..227K}. The importance of galactic foregrounds was highlighted by the claim of a detection of the primodial gravitational waves by the BICEP2 collaboration \citep{2014PhRvL.112x1101B} and retraction after incorporating the polarized emission from galactic dust \citep{2015PhRvL.114j1301B}, which had been underestimated \citep[e.g.,][]{2014JCAP...08..039F,2014JCAP...10..035M}. A better understanding of the ISM can thus aid in the interpretation of current and future microwave data. 

In addition, simulations of the microwave sky play an important role in the design of future CMB experiments. The most commonly used models in design studies are based on templates of synchrotron and dust emission derived from the \wmap{} and \planck{} data. Because future experiments are significantly more sensitive than \wmap{} and \planck{}, the data is smoothed to remove instrumental noise. This also removes the small scale galactic emission which is then reintroduced in the form of Gaussian random fluctuations \citep[e.g.,][]{2013A&A...553A..96D,2016MNRAS.462.2063H,2017MNRAS.469.2821T,2018MNRAS.476.1310M}. The ISM is highly non-Gaussian, and models derived from realistic MHD simulations that naturally capture the statistical properties of the ISM are likely to play a key role in future design studies and ultimately the analysis and interpretation of data \citep[e.g.,][]{2018JCAP...04..014D,2018arXiv180807445T}. 

Analytic theory and periodic box simulations allow to study the statistical properties applicable to the inertial range of turbulence. In addition, the scales of interest for future CMB missions for B-mode observations may include the energy injection scale (the reionization bump at $\ell \simlt 10$; \citealt{2016ARA&A..54..227K}). On large scales, the turbulence that shapes the distribution of the density and magnetic fields in the ISM is not the result of self-similar cascades, but is primarily driven by energy injection mechanisms and determined by interaction with large-scale physical structure. The predominant energy source of ISM turbulence is star formation feedback (mainly supernova explosions; \citealt{2004RvMP...76..125M,2007ARA&A..45..565M}). The star formation rates (SFRs) are highly time-dependent and spatially and temporally correlated. Therefore, comprehensive direct numerical simulations are necessary to produce realistic polarized signals from the multiphase, turbulent, magnetized ISM in galactic disks.

In this paper, we utilize a numerical simulation framework of the star-forming ISM called TIGRESS (Three-phase ISM in Galaxies Evolving with Resolved Star formation and Supernova feedback; \citealt{2017ApJ...846..133K}). In the TIGRESS simulation framework, we model local patches of vertically-stratified galactic disks, including self-consistent treatment of star formation and feedback, interstellar cooling and heating, magnetic fields, and galactic differential rotation. This local shearing box approach enables us to achieve uniformly high resolution to resolve key physics in the diffuse ISM. As a result, star formation rates, turbulence, magnetic fields, and thermal phases are self-consistently regulated by star formation feedback \citep{2017ApJ...846..133K}. Most importantly, in the context of dust polarization, the small-scale turbulent and large-scale mean magnetic fields are generated, grown, and saturated by supernova driven turbulence and galactic differential rotation \citep[e.g.,][]{2013MNRAS.430L..40G,2015ApJ...815...67K}.

In what follows, we construct a large number of synthetic dust polarization maps using a fiducial solar neighborhood model of the TIGRESS simulation suite \citep[][]{2017ApJ...846..133K,2018ApJ...853..173K}, analyze angular power spectra, and compare dust polarization characteristics observed in the simulations with the ISM properties observed by \planck{}. In Section~\ref{sec:maps}, we provide an overview of the numerical simulation (\S~\ref{sec:sim}) and a method to synthesize dust polarization maps (\S~\ref{sec:syn}). We then calculate power spectra as described in Section~\ref{sec:ps_method}, show example spectra in Section~\ref{sec:example}, and overall statistics in Section~\ref{sec:conv}. Section~\ref{sec:tevol} presents the full time evolution of the $EE/BB$ ratio and the $TE$  correlation seen in the synthetic maps. Sections~\ref{sec:tevol_local}, \ref{sec:tevol_qpo}, and \ref{sec:tevol_secular} respectively delineate the time evolution of dust polarization characteristics focusing on instantaneous, quasi-periodic, and secular variations in the context of the temporal fluctuations of the ISM properties. In Section~\ref{sec:summary}, we summarize the results and discuss  astrophysical and cosmological implications. Our synthetic maps are made publicly available\footnote{\url{http://tigress-web.princeton.edu/~changgoo/TIGRESS-dustpol-maps/index.html}}.

\section{Synthetic Dust Polarization Maps}\label{sec:maps}

\subsection{TIGRESS MHD simulation}\label{sec:sim}

Numerical MHD simulations that serve for polarized foreground studies for CMB experiments should provide realistic magnetized ISM structure as generated by nonlinear interactions between turbulence, sheared galactic rotation, self-consistent star-formation and supernova feedback. All of these elements are incorporated in the newly-developed TIGRESS framework \citep{2017ApJ...846..133K}, in which we model local patches of vertically-stratified galactic disks using the Athena MHD code \citep{2008ApJS..178..137S}. 

\begin{figure*}
\includegraphics[width=0.3\textwidth]{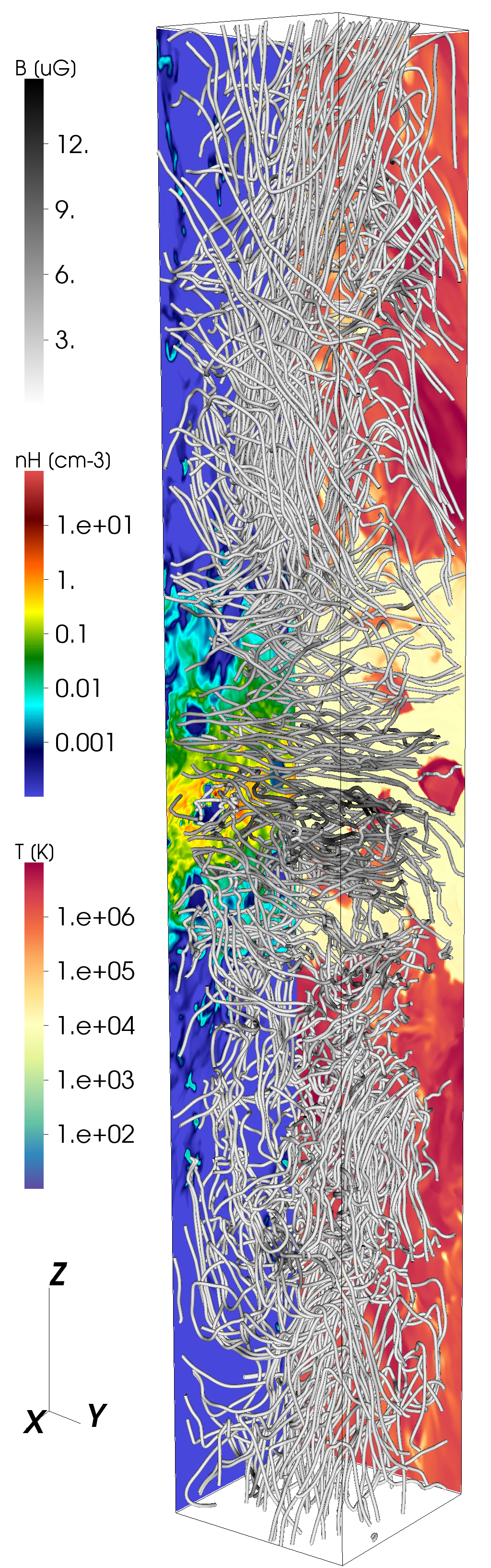}
\includegraphics[width=0.68\textwidth]{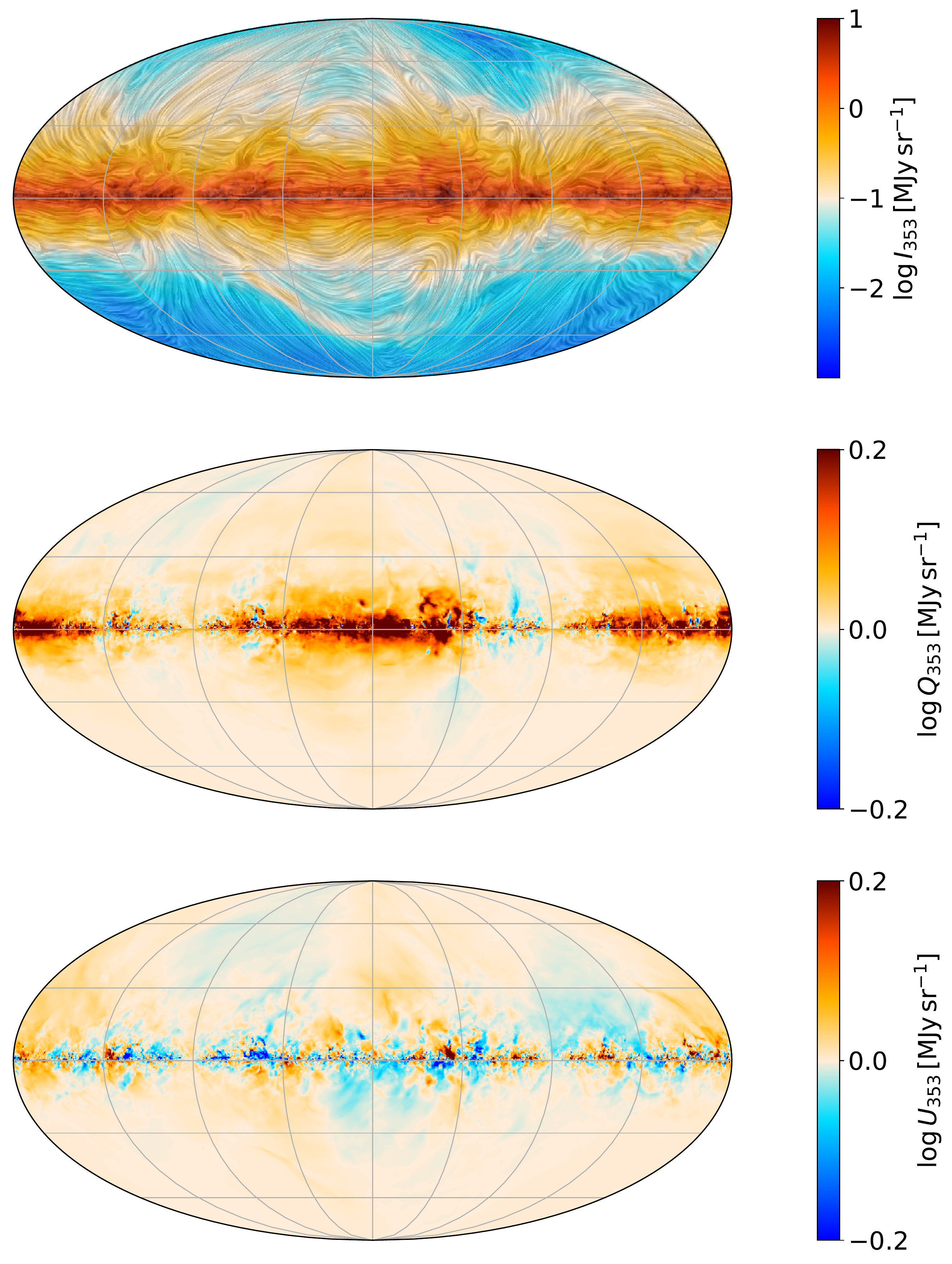}
\caption{\emph{Left}: Sample snapshot illustrating the magnetic field structure in a fiducial solar neighborhood model of TIGRESS at $t=360\Myr$.  Gas density and temperature slices are shown in the XZ and YZ planes, respectively. The outer dimension of the simulation is $1024\times1024\times7168 \pc^{3}$, and the spatial resolution is $\Delta x = 4\pc$, which is uniform throughout the simulation domain.
\emph{Right}: Stokes I, Q, and U parameters (from top to bottom) synthesized from the simulated density and magnetic fields shown in the left panel (see Section~\ref{sec:syn}).  The drapery patterns obtained by the line integral convolution are overlaid on the I map,  representing the apparent magnetic field orientation in the plane of the sky reconstructed from the polarization angle (rotated by $90^\circ$).
\label{fig:tigress}}
\end{figure*}

The advantage of this {\it shearing tall-box} setup is twofold. First, by modeling only a portion of galactic disks, we can achieve a uniformly high resolution to resolve key physics in the diffuse ISM. Our fiducial simulation adopts uniform spatial resolution $\Delta x =4\pc$ using a number of zones of $(N_x, N_y, N_z) = (256, 256, 1792)$ with a box size of $(L_x, L_y, L_z) = (1024, 1024, 7168)\pc$. Adaptive mesh refinement or (semi-)Lagrangian techniques like moving-mesh and smoothed particle hydrodynamics usually follow mass elements so that they under-resolve the diffuse ISM, which is crucial for the present purpose. Second, since it is global in the vertical direction, the realistic vertical large-scale structure of the ISM is well captured and allows a straightforward line-of-sight integration to construct reliable maps, especially toward the high-latitude sky. See Figure~\ref{fig:tigress} for a sample snapshot showing three dimensional magnetic field line structures along with slices of density and temperature.

The current TIGRESS framework includes: (1) star formation in self-gravitating gas with sink particles, (2) numerically well-resolved supernovae \citep{2015ApJ...802...99K} from star clusters and runaway OB stars, (3) photoelectric heating from a time-dependent FUV field, and (4) galactic differential rotation. \citet{2017ApJ...846..133K} have shown that a fiducial TIGRESS simulation with parameters similar to those of the solar neighborhood reaches a fully self-consistent quasi-steady state, with more than ten self-regulation cycles of star formation and feedback. The simulation shows realistic gas properties, including mass and volume fractions of cold, warm, and hot phases, turbulence velocities, magnetic field strengths (both mean and turbulent components), and disk scale heights. We refer the reader to \citet{2017ApJ...846..133K} for more complete descriptions of the TIGRESS framework. Since the ISM properties and SFRs in TIGRESS simulations are self-regulated for given galactic conditions, no specific parameter choice for turbulence and magnetic field strength and geometry is necessary.

\begin{figure}
\plotone{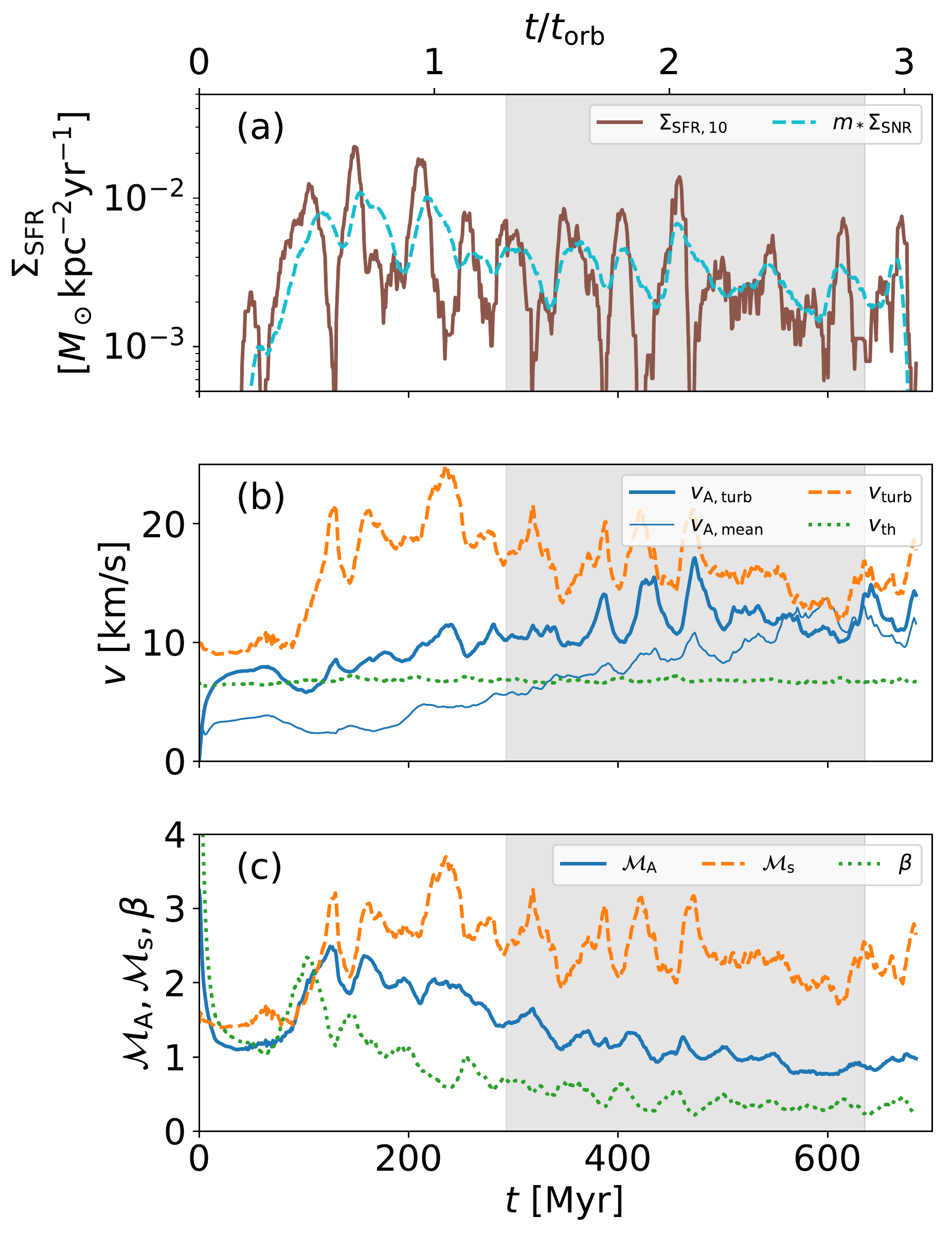}
\caption{Time evolution of (a) star formation and supernova rates, (b) mass-weighted turbulent ($\vturb$), Alfv\'en ($v_A$), and thermal ($c_s$) velocity dispersions and (c) sonic Mach number $\mathcal{M}_s$, Alfv\'enic Mach number $\mathcal{M}_A$, and plasma beta $\beta$ defined by Equations (\ref{eq:vturb})-(\ref{eq:beta}). The gray area shows the time range used to construct synthetic observations.
\label{fig:tevol}}
\end{figure}

To overview the key ISM properties in the simulation, Figure~\ref{fig:tevol} presents the time evolution of (a) surface density of star formation and supernova rates, (b) mass-weighted velocity dispersions, and (c) sonic and Alfv\'en Mach numbers and plasma $\beta$. The gray area is time range used to construct synthetic observations and analyze power spectra (with an interval of $\sim1\Myr$). The SFR surface density is calculated from the mass of young star cluster particles with the age of star cluster particles younger than $t_{\rm bin}=10\Myr$ (roughly equivalent to SFRs traced by H$\alpha$ emissions; \citealt{2012ARA&A..50..531K}). We count the total number of supernovae within the last $10\Myr$ to calculate the areal supernova rate $\Sigma_{\rm SNR}$ and plot in Figure~\ref{fig:tevol}(a) by multiplying the mass of young stars formed per SN, $m_*=95.5\Msun$ \citep{2017ApJ...846..133K}. The recent star formation and supernova rates are self-regulated to maintain an equilibrium state, $\Sigma_{\rm SFR,10}=2.4^{+1.8}_{-0.9}\times10^{-3}\sfrunit$ (ranges indicate the first and third quartiles), that is consistent with the observed $\Sigma_{\rm SFR}$ in the solar neighborhood \citep{2009AJ....137..266F}, with large temporal fluctuations.\footnote{Since our model does not have inflows, the gas surface density (and total gas mass) decreases gradually as the gas is gradually converted into the stars and escapes as galactic winds \citep{2018ApJ...853..173K}. The surface density during the time range of interest varies from $\Sigma=10.5\Surf$ to $8\Surf$, covering the solar neighborhood condition.}

The mass-weighted turbulent\footnote{For the turbulent velocity dispersion, we subtract the background shear arising from galactic differential rotation, $\vel_{\rm shear}=-q\Omega x\yhat$), and use the perturbed velocity of gas. $\delta \vel \equiv \vel-\vel_{\rm shear}$. Here, we adopt a flat rotation curve with the shear parameter $q\equiv|d\ln\Omega/d\ln R|=1$ and galactic rotational speed $\Omega=28\kms\kpc^{-1}$.}, Alfv\'en, and thermal velocity dispersions are respectively defined by
\begin{equation}\label{eq:vturb}
\vturb \equiv \rbrackets{\frac{\sum_{2p} \rho |\delta \vel|^2 dV}{\sum_{2p}\rho dV}}^{1/2},
\end{equation}
\begin{equation}\label{eq:va}
v_A \equiv \rbrackets{\frac{\sum_{2p} |\Bvec|^2/(4\pi) dV}{\sum_{2p} \rho dV}}^{1/2},
\end{equation}
\begin{equation}\label{eq:cs}
c_s \equiv \rbrackets{\frac{\sum_{2p} P dV}{\sum_{2p} \rho dV}}^{1/2}.
\end{equation}
Here, we further decompose Alfv\'en velocity into mean and turbulent components using the horizontally-averaged mean field $\overline{\Bvec}$ and turbulent field $\delta \Bvec \equiv \Bvec - \overline{\Bvec}$. To obtain representative values of velocity dispersions where most dust emission arises, we exclude the hot gas using a temperature criterion $T<2\times10^4\Kel$. We use the subscript ``2p'' in the summation to denote this ``two-phase'' medium (or the warm and cold medium). It is also useful to calculate dimensionless ratios between velocity dispersions, defining sonic and Alfv\'enic Mach numbers and plasma $\beta$ as
\begin{equation}\label{eq:machs}
\mathcal{M}_s \equiv \vturb/c_s,
\end{equation}
\begin{equation}\label{eq:macha}
\mathcal{M}_A \equiv \vturb/v_A,
\end{equation}
\begin{equation}\label{eq:beta}
\beta \equiv  \frac{\sum_{2p} P dV}{\sum_{2p} |\Bvec|^2/(8\pi) dV}= \frac{2c_s^2}{v_A^2}.
\end{equation}

We initially impose a uniform azimuthal magnetic field with a moderate strength ($\beta=10$). This gives an initial magnetic field at the midplane with $B_{y,0}(z=0)=2.3~\mu G$. As soon as turbulence is driven, the turbulent component of magnetic fields is rapidly created by tangling the initial fields. The turbulent Alfv\'en velocity dispersion is well correlated with the turbulent velocity dispersion with a slight offset as stronger turbulence makes magnetic fields more turbulent, while the mean Alfv\'en velocity smoothly grows through dynamo action \citep[e.g.,][]{2013MNRAS.430L..40G,2015ApJ...815...67K}. As the mean field slowly grows with the time scale comparable to the orbital period ($2\pi/\Omega =220\Myr$), the turbulent magnetic fields grows as well, and the total magnetic energy saturates at a level similar to the kinetic energy.

In our simulation, saturation of the turbulent fields occurs after $300\Myr$, while the mean field growth saturates after $500\Myr$ (see Figure~\ref{fig:tevol}(b)). We also find that the saturation level is independent of the initial field strength, but it takes longer to saturate with weaker initial field strength \citep[e.g.,][]{2015ApJ...815...67K}. At saturation, the strengths of the mean and turbulent magnetic fields become $|\overline{B}|=3.6~\mu G$ and $|\delta B|= 3.9~\mu G$, respectively. In the period of interest (the gray area in Figure~\ref{fig:tevol}), the secular evolution of magnetic fields enables us to cover the global gas properties $\mathcal{M}_A\sim1.6-1$ and $\beta\sim 0.7-0.2$, while the turbulent velocity fluctuations results in $\mathcal{M}_s\sim1.8-3.2$.

\subsection{Synthetic Dust Polarization Maps}\label{sec:syn}

Our simulation is carried out in local Cartesian coordinates. To synthesize all-sky dust polarization maps, we place observers in our simulation box. For a given observer's position, we tile the sky using {\tt HEALPix} \citep{2005ApJ...622..759G} with $N_{\rm side}=128$. The {\tt HEALPix} angles $(\phi, \theta)$ are the angles measured in the $x-y$ plane from $\xhat$-axis in the counterclockwise direction and from the $\zhat$-axis downward, respectively. At a given line-of-sight (LOS), the unit vectors along the LOS ($\Zhat$) and the plane of the sky ($\Xhat$ and $\Yhat$) are defined by rotations of the unit vectors of the Cartesian coordinates $(\xhat, \yhat, \zhat)$ by $\phi$ about $\zhat$-axis and $\theta$ about $\yhat$-axis (e.g., the {\tt HEALPix} Primer\footnote{\url{https://healpix.sourceforge.io/pdf/intro.pdf}} Fig. 5). The relationship between the two unit vectors are then given by
\begin{equation}\label{eq:coord_transform}
\begin{array}{llllllll}
\Xhat=&\cos \theta \cos \phi &\xhat &+& \cos \theta \sin \phi &\yhat &-&\sin \theta \zhat\\
\Yhat=&-\sin \phi &\xhat &+& \cos \phi &\yhat & &\\
\Zhat=&\sin \theta \cos \phi &\xhat &+& \sin \theta \sin \phi &\yhat &+& \cos \theta \zhat.
\end{array}
\end{equation}

Along each LOS, we sample the physical quantities from the observer's position to the maximum path length $\smax$ with an interval of $\Delta s \equiv \Delta x = 4\pc$. We apply the shearing-periodic boundary conditions \citep[e.g.,][]{1995ApJ...440..742H} whenever we march a ray outside the simulation domain horizontally. At every position along the LOS, we extract the gas density and the magnetic field vector from the original data cube via a trilinear interpolation using the eight nearest grid zones. At every snapshot, we construct nine realizations corresponding to nine observers at $x=-256, 0, 256\pc$ and $y=-256, 0, 256\pc$ in the $z=0$ plane.

Following the {\tt HEALPix} convention, the synthetic Stokes I, Q, and U parameters at frequency $\nu$ are  \citep[e.g.,][]{2015A&A...576A.105P}
\begin{eqnarray}
I_\nu&=&\int_{\smin}^{\smax} B_\nu(T_d)\left(1-p_0
\left(\frac{B_X^2+B_Y^2}{B^2} -\frac{2}{3}\right) \right)
\sigma_{d,\nu} n_H ds \label{eq:I}\\
Q_\nu&=&-\int_{\smin}^{\smax} p_0 B_\nu(T_d)\frac{B_X^2-B_Y^2}{B^2}
\sigma_{d,\nu} n_H ds \label{eq:Q}\\
U_\nu&=&-\int_{\smin}^{\smax} p_0 B_\nu(T_d)\frac{2B_XB_Y}{B^2}
\sigma_{d,\nu} n_H ds \label{eq:U}.
\end{eqnarray}
Here, $n_H=\rho/(1.4271 m_p)$ is the hydrogen number density, $B_X$ and $B_Y$ are the plane of the sky components of the magnetic field vectors projected by Equation (\ref{eq:coord_transform}), and the magnitude of the magnetic field is $B=(B_X^2 + B_Y^2+B_Z^2)^{1/2}$. Following \citet{2015A&A...576A.105P}, we adopt an intrinsic polarization fraction $p_0=0.2$, the blackbody source function $B_\nu$ with an uniform dust temperature $T_d=18\Kel$, and the dust opacity at 353~GHz $\sigma_{d,353}=1.2\times10^{-26}\cm^{-2}$.

Since our simulation utilizes uniform cubic grid zones, the coordinate conversion and integration along the radial directions cause non-trivial resolution effects. For example, due to the large covering area of nearby cells, angular resolution can become poor with an existence of nearby dense cells. If one aims to resolve modes up to a multipole moment $\lmax$, the minimum distance to the cell is
\begin{equation}\label{eq:smin}
\smin = \frac{\Delta x}{\pi}\lmax = 163\pc 
\rbrackets{\frac{\Delta x}{4\pc}}\rbrackets{\frac{\lmax}{128}}^{-1}.
\end{equation}
In practice, many observers are located in a real bubble in the simulation created by SN(e), reducing such contamination from nearby cells. For the sake of uniformity, however, we take $\smin=160\pc$, implying that we assume an artificial local bubble with radius of $160\pc$ surrounding an observer, similar to the Local Interstellar Bubble \citep[e.g.,][]{2018A&A...616A.132L}. We use $\smax = 3.5\kpc$ to cover a majority of the gas in the simulation. However, most of the signal comes from the gas within the disk scale height ($H\sim400\pc$) so that choosing any $\smax > 1\kpc$ results in similar synthetic maps. 

The right panel of Figure~\ref{fig:tigress} shows example Stokes I, Q, and U maps along with the drapery patterns calculated by the line integral convolution \citep{LICreference} representing the apparent magnetic field orientation in the plane of the sky. 

\section{Angular Power Spectra}\label{sec:PS}

\subsection{Method}\label{sec:ps_method}
We compute the angular power spectra of our synthetic maps using the nominal curved sky pseudo-$C_\ell$ method, which accounts for masks imposing sky cuts \citep[e.g.,][]{2002ApJ...567....2H,2003ApJS..148..161K,2005MNRAS.360.1262B}. This power spectrum estimator code has been thoroughly tested and was also used in analyzing the \wmap{} and \planck{} maps to study polarized galactic synchrotron and dust emission \citep{2015JCAP...12..020C}. We do not use the pure $B$-mode estimator in the following discussion as the synthetic maps have comparable power in $E$- and $B$-modes, but we have verified that the conclusions remain unchanged if a pure estimator is used.

We use $\ell$ bins of width 20 centered at 29.5, 49.5, 69.5, etc. The bin-to-bin correlation depends on the width of the $\ell$ bins, the sky fraction of the mask, and the smoothness of the mask apodization. To estimate the level of the bin-to-bin correlation, we generate 2000 Gaussian simulations with power law $EE$ and $BB$ spectra. For the mask with the smallest sky fraction described below, we find the average bin-to-bin correlation to be $< 5\%$. We compute $D_\ell$, where $D_\ell\equiv[\ell(\ell+1)/2\pi] C_\ell$.

\subsection{An Example}\label{sec:example}

\begin{figure}
\epsscale{0.8}
\plotone{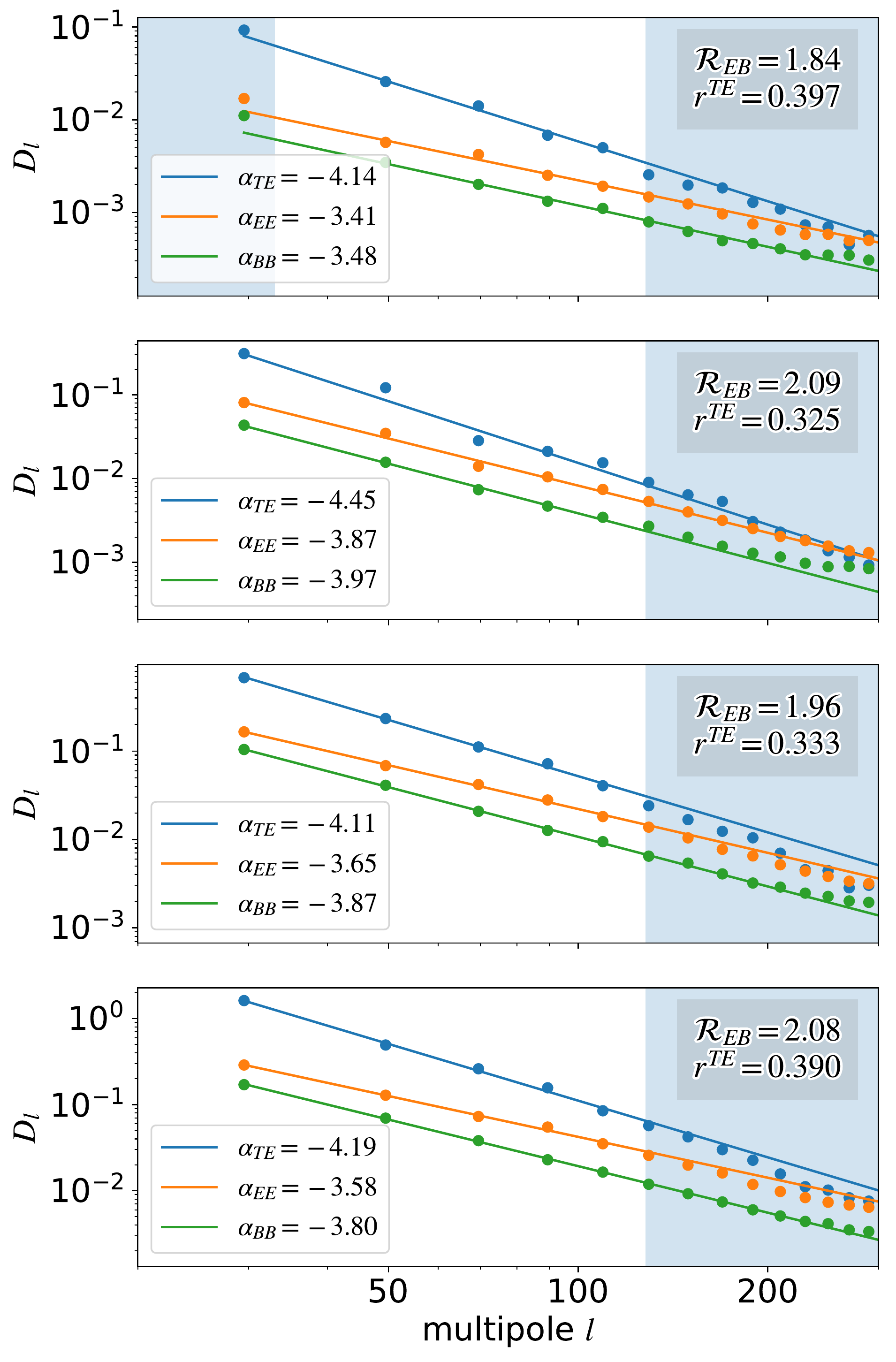}
\caption{Sample power spectra calculated by applying masks for $I_{\rm th} = 0.05$, 0.1, 0.2, and 0.3 MJy/sr (from top to bottom). The blue shaded regions are excluded for the power law fits. \label{fig:PS}}
\end{figure}

Figure~\ref{fig:PS} shows example $TE$ , $EE$, and $BB$ power spectra derived from synthetic maps presented in Figure~\ref{fig:tigress}. We construct masks for high dust intensity regions with four threshold values $I_{\rm th} = 0.05$, 0.1, 0.2, and 0.3 MJy/sr, which cover a range of the dust intensity adopted in PXXX. We also apply a latitude mask, masking $|b|<5.7^\circ$, where long path length with periodic boundary conditions would sample the same gas artificially multiple times. For this particular realization, from the higher to lower threshold value, the visible sky fractions are $\fsky=0.23$, 0.4, 0.56, and 0.66, respectively.

Following PXXX, we conduct least-square fits to the power spectra with a power-law function
\begin{equation}
D_\ell^{XY}=A_{80}^{XY} (80/\ell)^{\alpha_{XY}+2},
\end{equation}
where $XY$ can be $TE$ , $EE$, and $BB$. The fits use multipole ranges of $(\lmin, \lmax)$, where $\lmin \equiv 8 / \fsky$ and $\lmax=128$ (see Equation~\ref{eq:smin}).

We define the $EE/BB$ ratios at a given multipole and get the mean $EE/BB$ ratio (or the $E/B$ ratio, in short) $\REB\equiv \abrackets{D_\ell^{EE}/D_\ell^{BB}}$, where $\abrackets{}$ stands for the mean over the multipole range used in the power law fit. We also calculate the $E/B$ ratio using the amplitudes of the power-law fits, i.e., $A_{80}^{EE}/A_{80}^{BB}$, as in PXXX. The two values are consistent with each other, implying that the $EE/BB$ ratios are not strongly scale dependent; in general, $E$-mode power spectra have slightly shallower slopes than that of $B$-mode power spectra, although both power-law slopes are steeper in our simulations than those in \planck{} observations due to the projection effect (see Section~\ref{sec:conv}).

We also calculate the mean value of the dimensionless $TE$ correlation coefficient (or the $TE$ correlation, in short) at a given multipole $\rTE \equiv\abrackets{ D_\ell^{TE}/(D_\ell^{TT}D_\ell^{EE})^{1/2}}$. For the example realization shown in Figures~\ref{fig:tigress} and \ref{fig:PS}, the $E/B$ ratios and the $TE$  correlations are close to the \planck{} measurements, but, as we shall show later, this is a rare temporal case in the suite of synthetic maps.

\subsection{Overall Statistics and Convergence}\label{sec:conv}

We construct total 3,150 synthetic maps using 350 snapshots during $t\in(300,650)\Myr$ with $\Delta t=1\Myr$ for 9 observers in the midplane for each snapshot.\footnote{All synthetic maps are available at
\edit1{the author's private site, \url{http://tigress-web.princeton.edu/~changgoo/TIGRESS-dustpol-maps/index.html}, or the Legacy Archive for Microwave Background Data Analysis (LAMBDA) site, \url{https://lambda.gsfc.nasa.gov/simulation/tb_tigress_data.cfm}.}} We then calculate power spectra using masks with intensity thresholds ($I> 0.05$, 0.1, 0.2, and 0.3 MJy/sr) and latitude cuts ($|b|<5.7^\circ$) as described in the previous section. We further select the cases with $\fsky>0.16$ to ensure that the sky coverage is not too small to determine meaningful power spectra within the $\ell$ range of interest. We finally get 9,175 sets of power spectra, and at least 16 sets of power spectra at a given time. We refer to this result as {\tt fiducial} in what follows.

For comparison, we also synthesize 900 dust polarization maps for the first 100 snapshots using the warm medium only by applying a temperature cut, $5\times10^3 \Kel < T < 2\times10^4\Kel$, while for the calculation of power spectra, we apply the same masks used in the {\tt fiducial} maps. We refer to this result as {\tt warm-only}.

In addition, we create 900 maps using a lower resolution simulation ($\Delta x = 8pc$) for 100 snapshots during $t\in(300,400)$. We refer to this result as {\tt low-res}. Note that overall evolution and ISM properties are \emph{statistically} converged at this resolution \citep{2017ApJ...846..133K}, but this does not imply exactly the same ISM realization with lower resolution because the evolution of the star-forming ISM is chaotic in nature. Therefore, the resolution comparison should be understood in a \emph{statistical} sense.

\begin{figure}
\epsscale{1.0}
\plotone{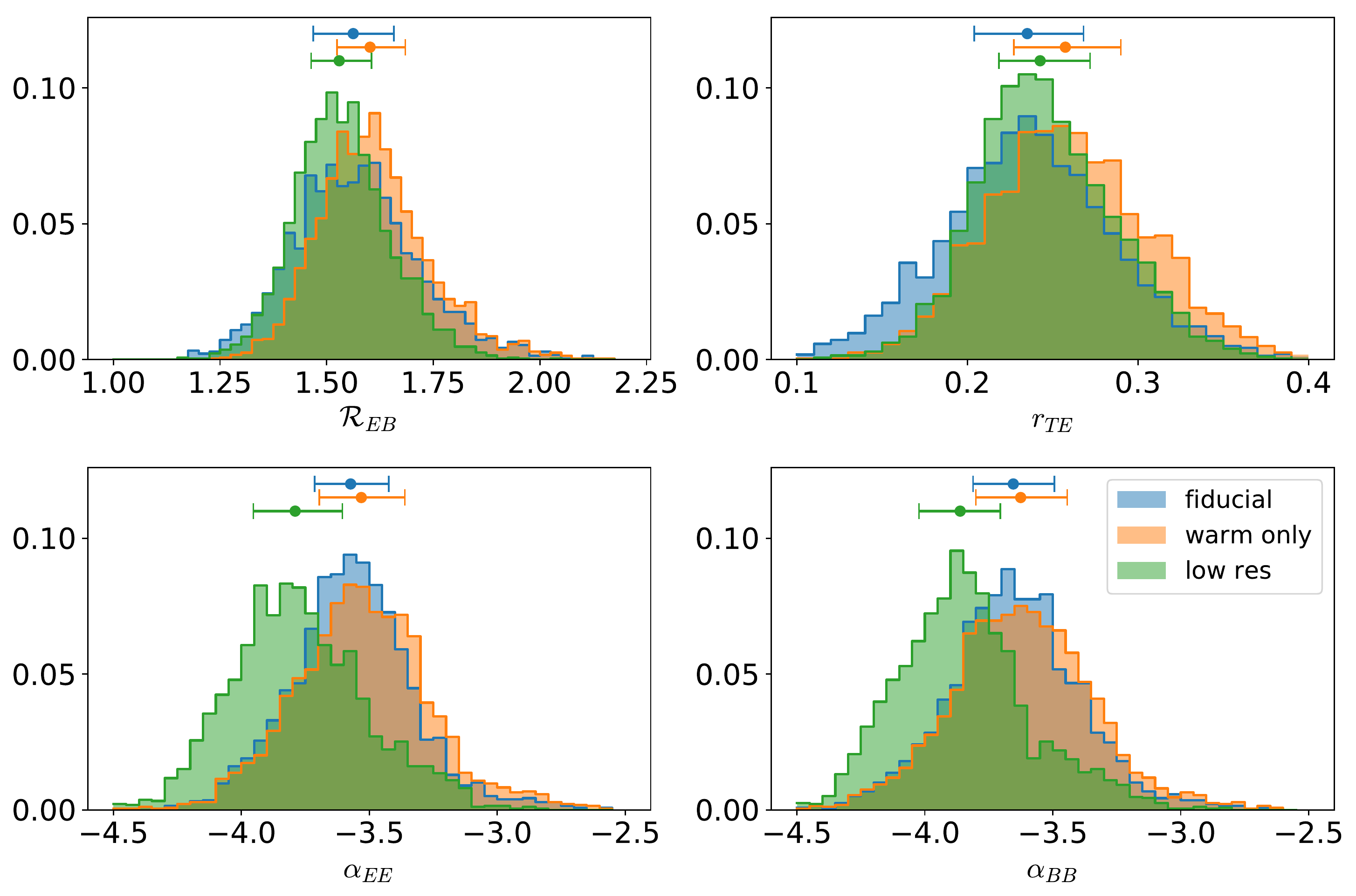}
\caption{Distribution of fitting parameters for the {\tt fiducial} (blue), {\tt warm-only} (orange), and {\tt low-res} maps (green). The {\tt warm-only} maps are constructed by using the fiducial simulation, but selecting the warm medium ($5\times10^3 \Kel < T < 2\times10^4\Kel$) only, and the {\tt low-res} maps are constructed by using the low-resolution simulation. The circle on top of the histogram gives the median, while the errorbars enclose the first and third quartiles. There are no apparent significant systematic differences in the values of $\REB$ and $\rTE$ among three sets of maps, while the {\tt low-res} maps produce systematically steeper slopes than the others. 
\label{fig:conv}}
\end{figure}

Figure~\ref{fig:conv} shows the overall statistics of the $E/B$ ratios and $TE$ correlations as well as power law slopes of $EE$ and $BB$ power spectra. Note that we only show results from the {\tt fiducial} maps from the first 100 snapshots to enable direct comparison with the {\tt warm-only} and the {\tt low-res} maps.

As noted in Figure~\ref{fig:PS}, the slopes of the power spectra are generally steeper than those in the \planck{} observation. This is largely caused by the limited coverage of the inertial range of turbulence in the simulation in combination with the sky projection. \edit1{The power spectra in turbulence are characterized by an inertial range with a well-defined power-law (either due to a self-similar energy transfer \citep{1941DoSSR..30..301K} or a collection of shocks \citep{1972LNP....12...41B})} and a dissipation range with a steep drop of power \citep[e.g.,][for an inviscid, isothermal, compressible turbulence simulation]{2003ApJ...590..858V}. In inviscid numerical simulations (i.e., without explicit viscosity and resistivity), the dissipation range is set by numerical schemes and resolution, typically happening at scales of $\sim 10-20$ resolution elements depending on the numerics. \edit1{In our simulations, we also find that the kinetic and magnetic energy spectra shows a change of slopes near the wavenumber at $kL_{x,y}/(2\pi) \sim 20-30$ or the linear length scale at $\lambda\sim 30-50\pc$.\footnote{\edit1{We defer the detailed analysis and discussion of the turbulence properties of TIGRESS simulations in the future work. In order to get a general idea of the inertial range, we obtain kinetic and magnetic energy spectra as a function of the horizontal wavenumbers $k_x$ and $k_y$ from slices at different height $z$. We take time average over $t=300-400\Myr$ to enhance statistical significance. Both energy spectra show a smooth turnover of slopes at the wavenumber of $20-30$ rather than a sharp break. The magnetic energy spectra show shallower slopes than the kinetic energy spectra with gradual steepening toward the midplane. The kinetic energy spectra are consistent to $k^{-2}$ at all heights.}}}

For projections along the Cartesian axes, there is a one-to-one correspondence between wavenumber and multipole moment, and the inertial and dissipation ranges are well separated. As a consequence, one can infer the slope of power spectra even for a relatively narrow inertial range \citep[e.g.,][]{2018PhRvL.121b1104K}. For the radial projection of the simulations onto the sky used here, there is no correspondence between wavenumber and angular scale because even small nearby structures appear large in the sky. This artificially steepens the power-law slopes significantly compared to their values in the inertial range.

To investigate the effect more closely, we have also considered projections along the Cartesian axes. We take Equations (\ref{eq:I})-(\ref{eq:U}), but consider the vertical axis ($\zhat$) instead of the radial axis ($\Zhat$). We integrate toward the upper boundary $z_{\rm max}=L_z/2$ by varying $z_{\rm min}=0$, 128, $\cdots$, 1024~pc. The $E/B$ ratio and slopes in the $EE$ and $BB$ power spectra are essentially unchanged if $z_{\rm min}>128\pc$ (equivalent to the intensity masks to exclude the cold, dense medium near the midplane). 

We find shallower slopes ranging between \mbox{$-2.9<\alpha<-2.6$} significantly closer to the slopes observed in \citet{2018PhRvL.121b1104K} and the \planck{} data. In addition, since \citet{2018PhRvL.121b1104K} observe an $E/B$ ratio that approaches unity on small scales where numerical dissipation is significant, one may also expect a bias on the $E/B$ ratio and $TE$  correlation in the radial projections. However, we find that the observed $E/B$ ratio and $TE$  correlation coefficients for the projections along the Cartesian coordinates are similar to those obtained for the radial projections. 

This is further supported by comparisons between {\tt fiducial} (blue) and {\tt low-res} (green) results shown in Figure~\ref{fig:conv} which demonstrate that the systematic biases in the measurements of $\REB$ and $\rTE$ are less problematic, while the steepening of power spectra at lower resolution is evident in both $EE$ and $BB$ power spectra.

The {\tt warm-only} maps also show consistent values of $\REB$ and $\rTE$ (slightly higher than the {\tt fiducial} maps), while the slopes are the same. As we already masked out low-latitude, high-intensity skies, our {\tt fiducial} maps are mostly dominated by the warm medium, which is better resolved than the cold medium. Overall, we conclude that our synthetic maps are converged in the sense that the measured $E/B$ ratio and $TE$  correlation coefficient are not sensitive to the projection and resolution, allowing us to investigate these two polarization characteristics in more detail. 

The effects of the radial projection on the spectral slope can be mitigated by increasing the resolution near the observer, but we leave this for future work.

\section{Time Evolution of Polarization Characteristics}\label{sec:tevol}

Our numerical simulation provides a long-term evolution of the ISM for conditions similar to those in the solar neighborhood, with self-consistent changes of turbulence energy injection due to self-regulation of SFRs as well as corresponding changes of the ISM state (see Figure~\ref{fig:tevol}). Therefore, it is interesting to explore time evolution of the dust polarization characteristics rather than analyzing time averaged ``stacked'' power spectra.

\begin{figure}
\epsscale{1.0}
\plotone{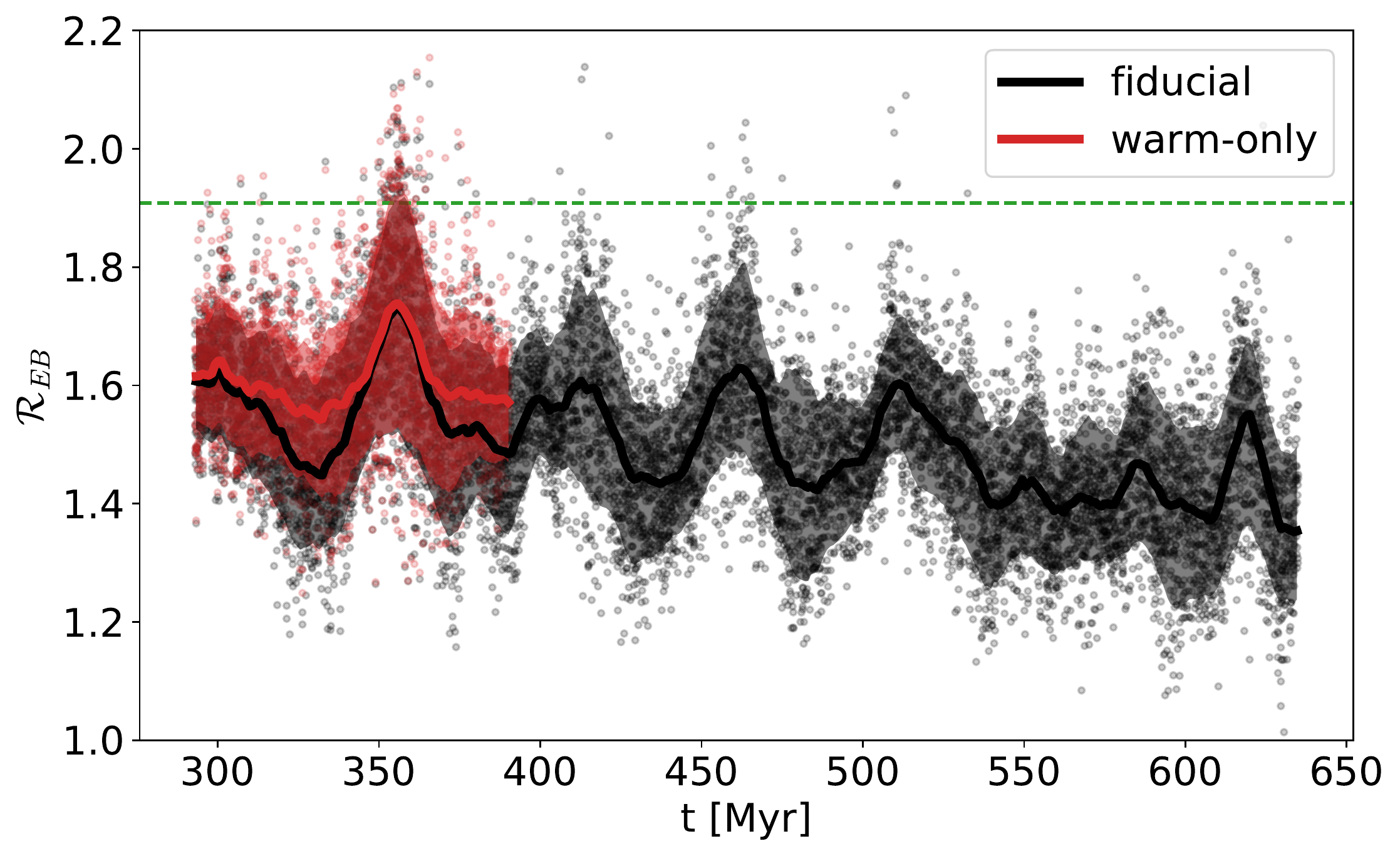}
\plotone{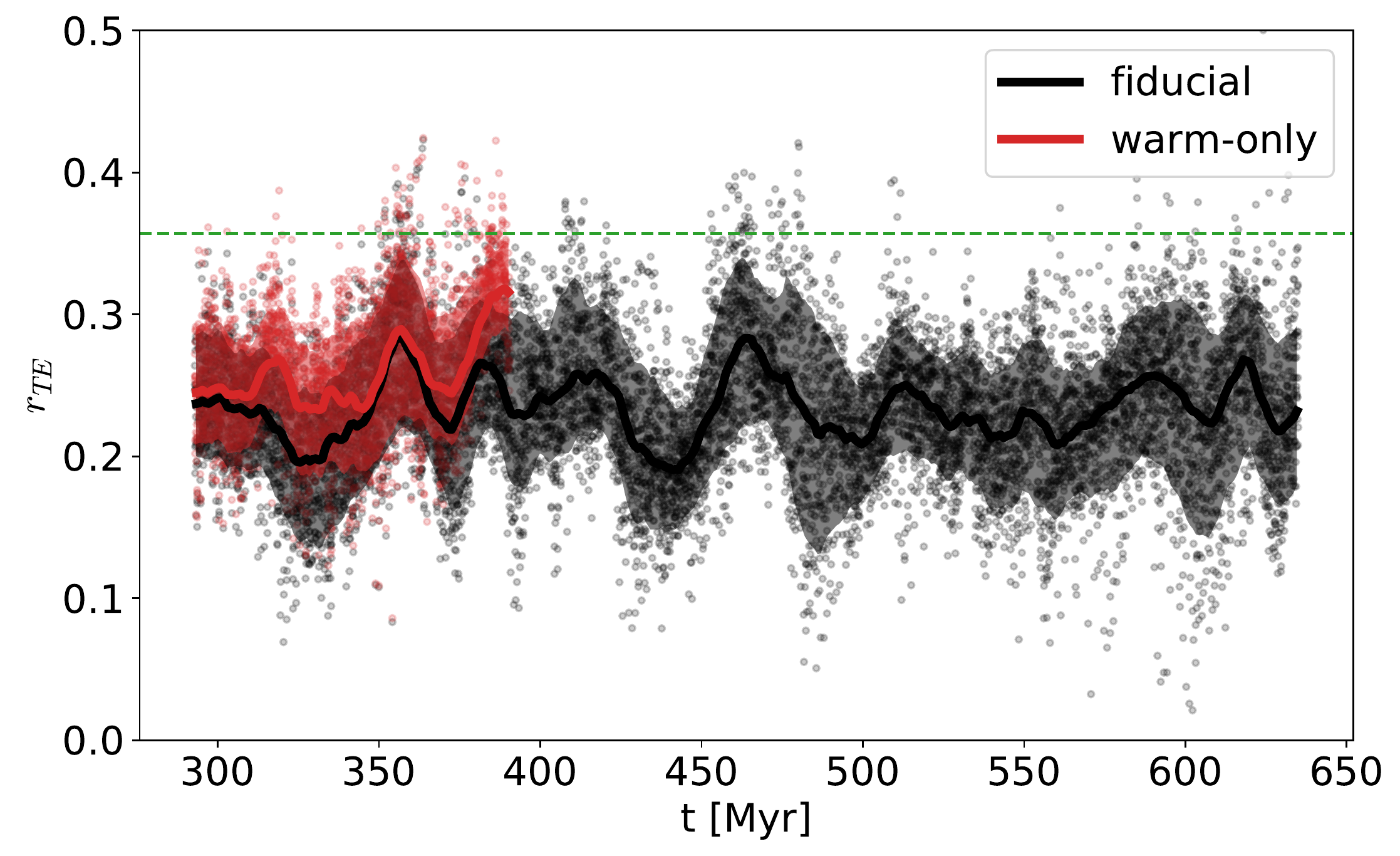}
\caption{$E/B$ ratios $\REB$ (top) and $TE$  correlations $\rTE$ (bottom) for all power spectra as a function of time. The horizontal dashed lines indicate the $E/B$ ratio and the $TE$  correlation reported in PXXX, $\REB=1.91$ and $\rTE=0.357$. The lines in both panels present the rolling mean with the time window of $10\Myr$ for the median values at a given time, while the shaded regions enclose between 10th and 90th percentiles. The {\tt warm-only} maps are shown in red. \label{fig:tevol_EB_TE}}
\end{figure}

Figure~\ref{fig:tevol_EB_TE} shows $E/B$ ratios $\REB$ and $TE$ correlations $\rTE$ for all power spectra as a function of time. Note that we have a maximum of 36 (9 observers and 4 masks) power spectra at a given time, but the number can be as small as 16 after discarding cases with low sky fraction ($\fsky<0.16$). The median and 10th and 90th percentiles smoothed by the rolling mean with the time window of $10\Myr$ are shown as the solid line and shaded region, respectively. The results from the {\tt warm-only} maps are shown as red in the same figure.

\begin{figure}
\epsscale{1.0}
\plotone{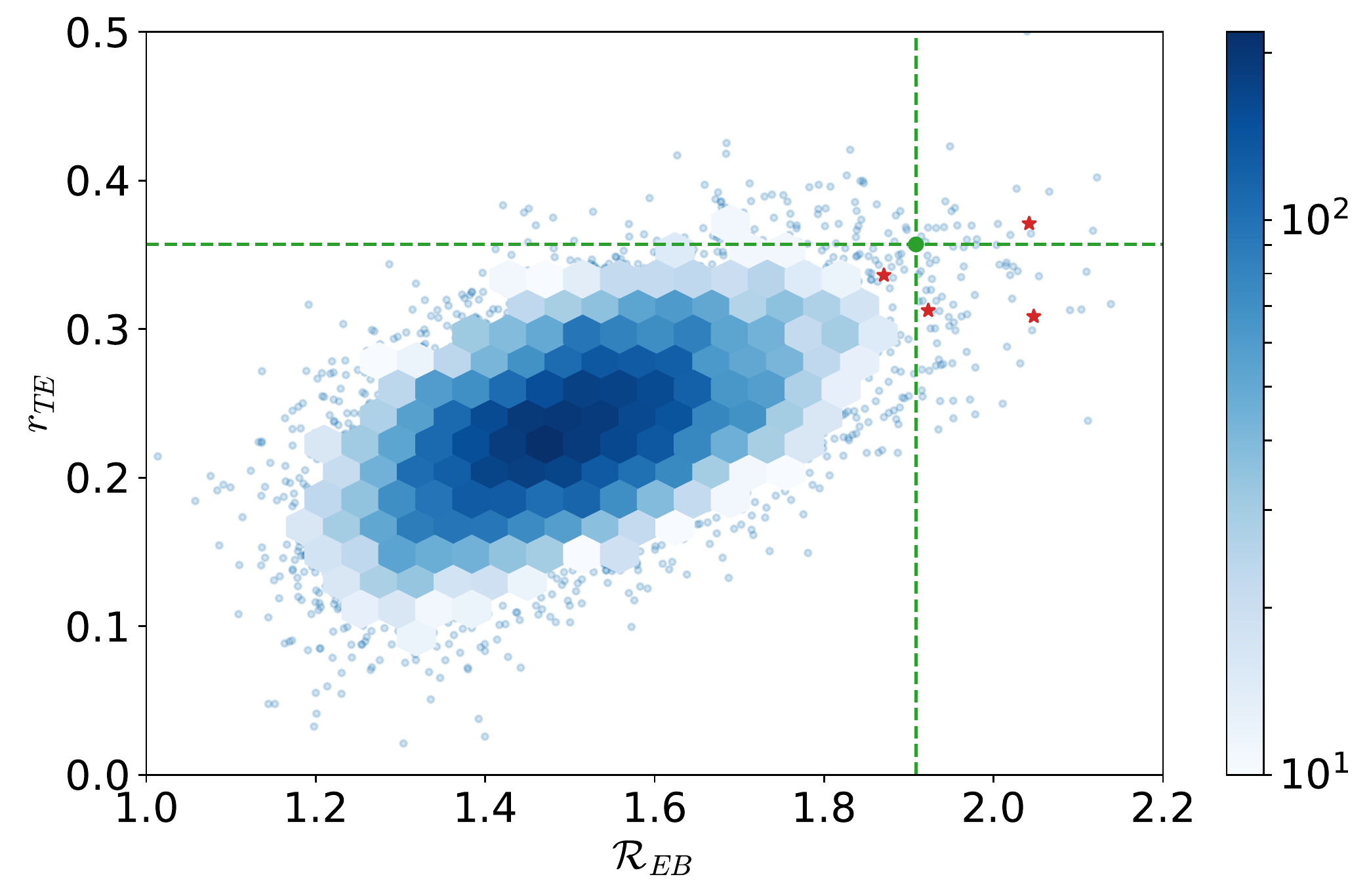}
\caption{Heat map between $\REB$ and $\rTE$. The \planck{} measurements are shown as dashed lines. The values from the example power spectra in Figure~\ref{fig:PS} are shown as stars. \label{fig:EB_TE}}
\end{figure}

First of all, we emphasize that we find ubiquitous $E/B$ power asymmetry and positive $TE$ correlation in the synthetic maps as seen by PXXX. However, the median values of $\REB\sim1.5$ and $\rTE\sim0.23$ are smaller than those measured by \planck{}. Figure~\ref{fig:EB_TE} presents a heat map between $\REB$ and $\rTE$, showing that $\REB$ and $\rTE$ are generally correlated.

Given the generally higher $E$-mode power than $B$-mode power and positive $TE$  correlation, in what follows, we focus on temporal fluctuations shown in the synthetic maps. There are three main features in the time evolution of $\REB$ and $\rTE$ that we separately address in the following subsections: (1) large instantaneous variations, (2) quasi-periodic fluctuations, and (3) secular evolution.

\subsection{Large Instantaneous Variations}\label{sec:tevol_local}

We observe large variations in both measured characteristics $\REB$ and $\rTE$ from the same snapshot depending on observer's positions. The mean levels of fractional variations within the same snapshot are $\delta \REB = 0.18$ and $\delta \rTE =0.45$, where $\delta q\equiv(q_{90} - q_{10})/q_{50}$ is defined using the $n$-th percentile of a quantity $q$, $q_{n}$. Note that the levels of fractional variations arising from different masks (intensity thresholds) are $\delta \REB = 0.085$ and $\delta \rTE = 0.24$, about a half of the total fractional variation levels.

It is natural to expect that synthetic observations from the same snapshot would share similar dust polarization characteristics since they are looking through the same ISM state. However, the dust polarization signals mainly come from the nearby medium (density-weighted). In other words, the scale length of signal should be comparable to the (projected) scale height of the warm/cold medium ($H\sim 400\pc$ in the simulation; \citealt{2017ApJ...846..133K}). The energy injection scale of turbulence is also similar to the scale height of the disk. This means that the synthetic skies seen by different observers separated by a few hundreds of parsec are effectively independent realizations.

\subsection{Quasi-periodic Fluctuations}\label{sec:tevol_qpo}

In addition to large instantaneous variations, Figure~\ref{fig:tevol_EB_TE} also shows quasi-periodic fluctuations of mean values of both $\REB$ and $\rTE$, similarly with ISM properties presented in Figure~\ref{fig:tevol}. The period of these quasi-periodic fluctuations is roughly $\sim 40-50\Myr$, corresponding to the period of the duty cycle of star formation self-regulation; when SFRs are enhanced, SN feedback expels gas and destroys dense clouds near the midplane, and star formation is quenched until the expelled gas falls back and is recollected by gravity and large scale flows into dense clouds. Here, the governing time scale of this process is the vertical dynamical time scale, \mbox{$t_{\rm ver}\sim 2H/v_z\sim 40-50\Myr$} \citep{2013ApJ...776....1K,2015ApJ...815...67K,2017ApJ...846..133K}.\footnote{ The star formation duty cycle is also related to the star cluster lifetime (or duration of feedback); since SN rates from one star cluster are slowly decreasing as a function of time, one burst can persist longer than the vertical dynamical time. Governing time scale for the star formation duty cycle is longer one between the vertical dynamical time and feedback lifetime. For our adopted population synthesis model ({\tt Starburst99}; \citealt{1999ApJS..123....3L}), the half-life of SN rates is about $20\Myr$ \citep[see][]{2017ApJ...846..133K}.}

\begin{figure}
\plotone{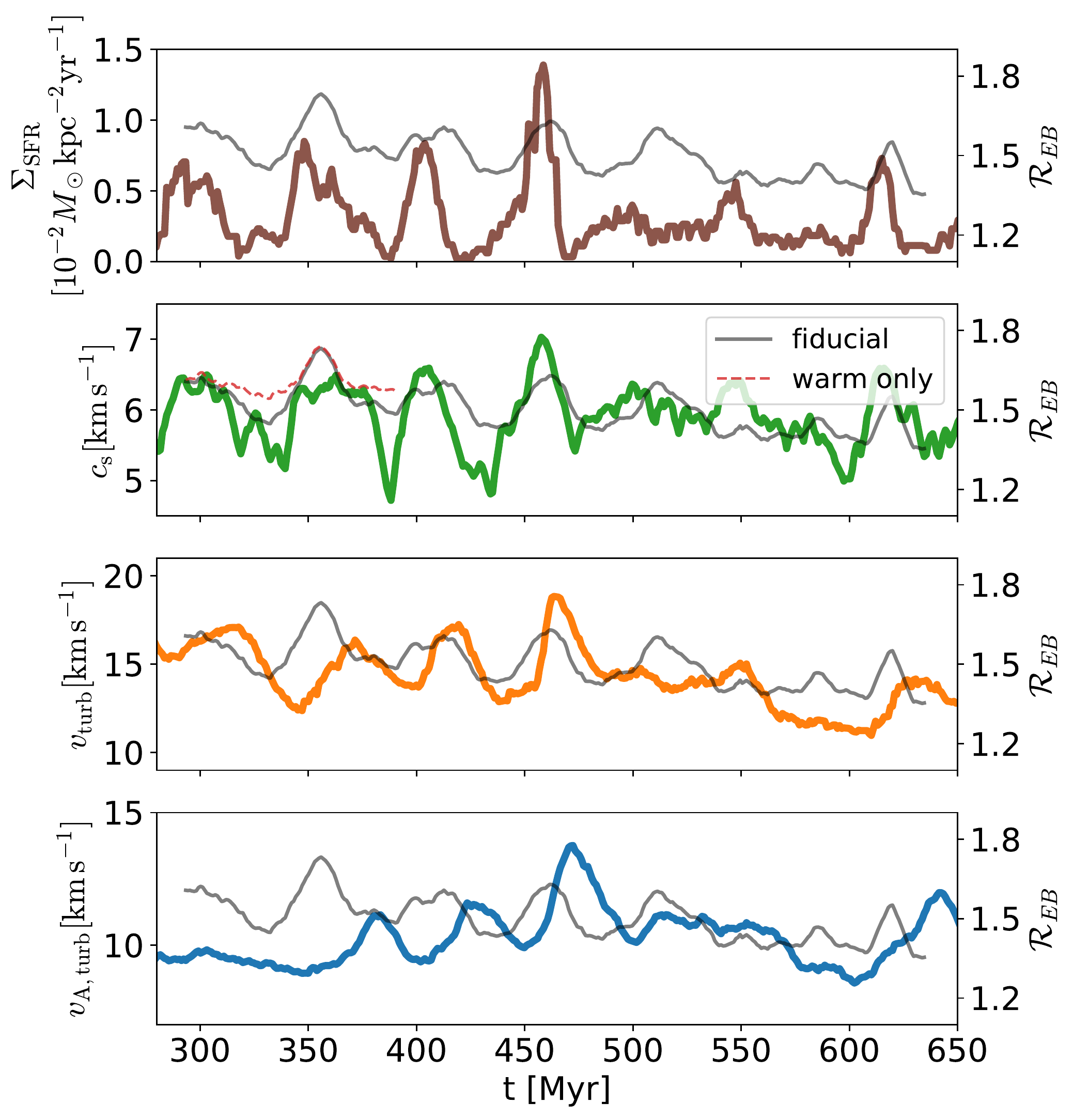}
\caption{From top to bottom, time evolution of SFR surface density $\Sigma_{\rm SFR}$, mass-weighted sound speed $c_s$, turbulent velocity dispersion $v_{\rm turb}$, and turbulent Alfv\'en velocity $v_{\rm A,turb}$, along with the $E/B$ ratio $\REB$ (on the right y-axis). Quasi-periodic temporal fluctuations are evident in all quantities with apparent correlation between $c_s$ and $\REB$. \label{fig:tevol_comp}}
\end{figure}

The duty cycle of self-regulation also results in offsets between SFRs and thermal, turbulent, and magnetic components of velocity dispersion (or pressure). To show this clearer, Figure~\ref{fig:tevol_comp} plots SFR surface density and velocity dispersions shown in Figure~\ref{fig:tevol} as a function of time along with the median values of $\REB$ shown in Figure~\ref{fig:tevol_EB_TE}, with an emphasis on fluctuations by adjusting plotting ranges. From top to bottom, we compare $\REB$ with SFR surface density, sound speed, turbulent velocity, and turbulent Alfv\'en velocity. One can clearly see that the patterns in velocity dispersions are shifted progressively from $c_s$ to $v_{\rm turb}$ to $v_{\rm A, turb}$ as each component is controlled by different physical processes of star formation feedback \citep[see][]{2010ApJ...721..975O,2011ApJ...743...25K,2011ApJ...731...41O,2013ApJ...776....1K,2015ApJ...815...67K}. 

If SFRs are higher than their equilibrium values, the warm medium gas fraction ($f_w$) increases almost immediately since FUV radiation provided by massive stars younger than $\sim 10\Myr$ enhances the photoelectric heating rate from small grains, which is a dominant heating process in the diffuse ISM \citep[e.g.,][]{1995ApJ...443..152W}. The mass-weighted mean sound speed $c_s\approx f_w c_w$ \citep[e.g.,][]{2010ApJ...721..975O} thus simply follows the recent SFRs. After $\sim 4-5\Myr$ from the birth of star clusters, massive stars begin to die and explode as SNe, driving turbulence. The turbulent velocity dispersion peaks after $\sim 10\Myr$ from the peak of $\SigmaSFR$ and $c_s$. Then, turbulent magnetic fields are generated by tangling of existing magnetic fields due to turbulence. The peak of the turbulent Alfv\'en velocity is expected to be further delayed after reaching the peak of turbulence.

At a glance, $\REB$ shows a positive correlation with $\SigmaSFR$ and $c_s$, especially for $t<500\Myr$. \citet{2018PhRvL.121b1104K} pointed out that dense gas, which is masked out with a number density threshold $n_{\rm th} = 70 \pcc$ in their synthetic dust polarization maps before projection, can randomize polarization to significantly reduce $\REB$. When $c_s$ is lower and the cold medium fraction is higher, the intervening cold medium is more likely to reduce the overall $\REB$.

A comparison between $\REB$ from the {\tt fiducial} and {\tt warm-only} maps can provide evidence for the general agreement with this expectation (see the second panel of Figure~\ref{fig:tevol_comp}). $\REB$ in the {\tt warm-only} maps are larger than that in the {\tt fiducial} maps (Figure~\ref{fig:conv} also shows slight increase of $\REB$ for the warm medium only case), especially when $\REB$ is low (in other words, when the cold medium dominates). Therefore, the dips in $\REB$ are less prominent in the {\tt warm-only} maps, in which less randomization effects are expected by the cold medium. However, at the peak of $\REB$, there is almost no difference between {\tt fiducial} and {\tt warm-only} maps as the volume is already dominated by the warm medium. 

While reduced, the {\tt warm-only} maps still show similar temporal fluctuations. Simple considerations for the variation of the warm medium fraction alone cannot fully explain the quasi-periodic oscillations in $\REB$. Although there is another apparent (anti-)correlation between $\REB$ and the turbulent Alfv\'en velocity, we do not push this analysis further with just simple, volume-integrated turbulence properties. A more careful analysis on characteristics of MHD turbulence in compressible, multiphase ISM driven by SN feedback like our simulation is needed to show any causal relations between SFRs, ISM properties, and dust polarization characteristics more robustly. 

\begin{figure}[th]
\plotone{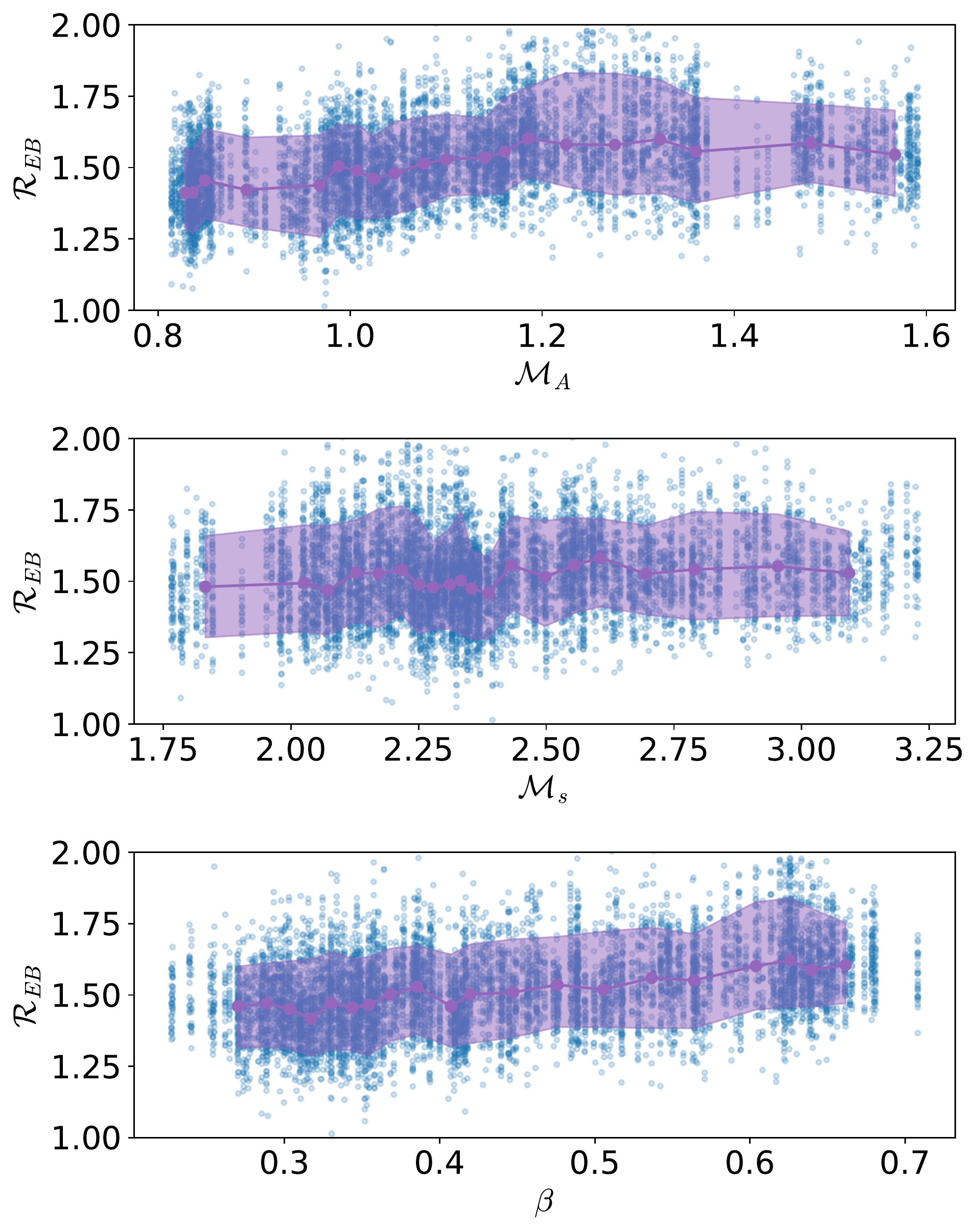}
\caption{Correlation of the $E/B$ ratio $\REB$ with Alfv\'en Mach number $\mathcal{M}_A$ (top), sonic Mach number $\mathcal{M}_s$ (middle), and plasma beta $\beta$ (bottom). The circles are median values of $\REB$ binned by the sorted $x$-axis quantity with 5\% intervals so that each point has the same statistical significance. The shaded region encloses 10 to 90 percentiles of $\REB$ within bins.
\label{fig:EB_comp}}
\end{figure}

\subsection{Secular Evolution}\label{sec:tevol_secular}

In the simulation, the total gas mass is decreasing due to star formation and mass loss, and the mean magnetic fields are growing due to galactic dynamo (Figure~\ref{fig:tevol}). They drive the secular evolution of SFRs and the ISM state, which potentially drive the secular evolution of the dust characteristics as well. Although local and temporal variations are much more prominent, there is a weak secular decreasing trend in $\REB$ but not in $\rTE$ (Figure~\ref{fig:tevol_EB_TE}).

\begin{figure}[th]
\plotone{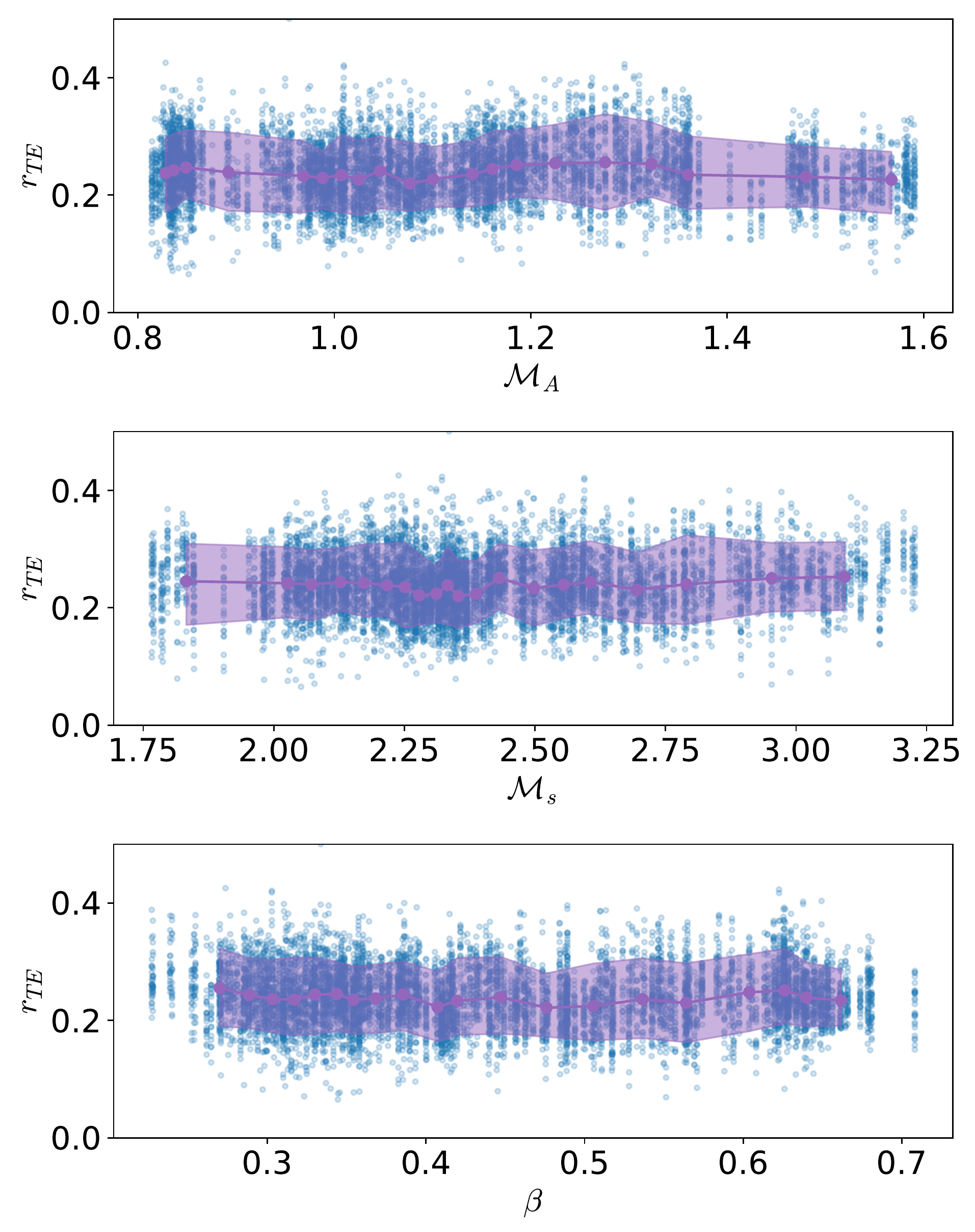}
\caption{Same as Figure~\ref{fig:EB_comp} but for the $TE$  correlation coefficient $\rTE$. \label{fig:TE_comp}}
\end{figure}

To gauge the relative importance between thermal, turbulent, and magnetic states of the ISM for the dust characteristics, Figures~\ref{fig:EB_comp} and \ref{fig:TE_comp} respectively plot $\REB$ and $\rTE$ measurements as a function of the dimensionless parameters such as Alfv\'en Mach number $\mathcal{M}_A$ (Eq.(\ref{eq:macha}); top), sonic Mach number $\mathcal{M}_s$ (Eq.(\ref{eq:machs}); middle), and plasma $\beta$ (Eq.(\ref{eq:beta}); bottom). Overall, $\REB$ and $\rTE$ are not strongly correlated with these \emph{global} ISM parameters.

While less distinctive, $\REB$ increases with both $\mathcal{M}_A$ and $\beta$, implying that $\REB$ is generally higher with weaker magnetic fields. This is consistent with the apparent anti-correlation between $\REB$ and turbulent Alv\'en velocity seen in Figure~\ref{fig:tevol_comp}. It is also possible that $\REB$ is just higher with stronger turbulence (may relate to higher SFRs).

Intriguingly, $\rTE$ does not show such a correlation due to the secular evolution although there is a general correlation between $\REB$ and $\rTE$ as shown in Figure~\ref{fig:EB_TE}. Since similar temporal fluctuations are shown in Figure~\ref{fig:tevol_EB_TE} for both $\REB$ and $\rTE$, the correlation between $\REB$ and $\rTE$ should arise from short-term variation of the ISM state that is responsible for quasi-periodic fluctuations rather than secular variation of turbulence properties. 

In addition, comparing the state of the ISM at early and late times in the simulation (Figure~\ref{fig:tevol}), we see that turbulence gets weaker (both kinetic and magnetic components) as the mean field grows, but the sound speed is unchanged. The lack of secular variation of $\rTE$ with still substantial quasi-periodic fluctuations thus implies that $\rTE$ may be a more sensitive probe of the thermal state of the ISM than of the turbulence state, while both ISM properties can be responsible for $\REB$. 

\vskip 1.5cm
\section{Summary \& Discussion}\label{sec:summary}

We have presented the first set of full-sky synthetic dust polarization maps constructed from a self-consistent MHD simulation. Our simulation utilizes the TIGRESS framework \citep{2017ApJ...846..133K}, in which the magnetized ISM in a local patch of galactic disks evolves self-consistently under the influence of interstellar cooling and heating, galactic differential rotation, self and external gravity, and turbulence driven by SN feedback. We have constructed a large number of synthetic maps seen by nine representative observers in the galactic plane for time snapshots covering $\sim 350\Myr$ ($\sim 1.5$ orbital time; 6-7 feedback cycles). Our analysis of these maps demonstrates that the recent \planck{} observations of $E/B$ power asymmetry and positive $TE$ correlation emerge naturally in MHD simulations of the star-forming ISM. Our main results are summarized in more detail as follows.

\begin{enumerate}
\item From our large number of synthetic skies (3,150), we find that the $EE/BB$ ratios ($\REB$) are larger than unity and that the dimensionless $TE$  correlation coefficients ($\rTE$) are positive, in broad agreement with the \planck{} observations. However, the typical values seen in our synthetic skies are $\REB\sim 1.4-1.7$ and $\rTE\sim 0.2-0.3$, somewhat lower than the values reported by \planck{}, except for short periods during which the \planck{} values are reached.

\item  Numerical simulations are inherently limited in the range of scales that are resolved, and numerical dissipation occurs at scales near the grid resolution, where we expect a steep drop of turbulence power. The radial projection onto the sky necessary to generate the full-sky maps mixes up the inertial and dissipation ranges of turbulence so that the angular power spectra for our synthetic skies are affected by numerical dissipation. This results in steeper slopes (see Figure~\ref{fig:PS}) than in \citet{2018PhRvL.121b1104K} and PXXX. To quantify this effect, we also consider a projection along the Cartesian axes and find that the slopes measured in the inertial range are significantly closer to those in both \citet{2018PhRvL.121b1104K} and PXXX. However, the values for the $E/B$ ratio and $TE$ correlation we obtain for the projections along the Cartesian axes are consistent with those obtained for the radial projection and are somewhat lower than those observed by \planck{}. This gives us confidence in the measurements of these two characteristics from our simulations, although we cannot fully exclude the possibility that numerical dissipation and imperfect separation of the dissipation range introduces some bias. To eliminate the mixing of scales as a possible bias, the simulations should be further developed to increase the resolution near the observers. Higher resolution simulations are generally also required to better resolve the diffuse cold neutral medium and to better understand whether values of the $E/B$ ratio and the $TE$  correlations seen by \planck{} are indeed rare. We leave this for future work.

\item The dynamic ISM model, which includes self-consistent time evolution due to self-regulation of star formation rates, shows a distinctive temporal evolution of both $\REB$ and $\rTE$. $\REB$ shows a quasi-periodic temporal fluctuations with period similar to the star formation duty cycle and a secular decreasing trend following the growth of the mean fields and/or decreasing SFRs. $\REB$ and $\rTE$ are generally correlated, but $\rTE$ does not show the secular decrease.

\item \citet{2017ApJ...839...91C} and \citet{2017MNRAS.472L..10K} have provided methods to calculate polarization power spectra for dust emission from MHD turbulence theory (see also \citealt{2018MNRAS.478..530K} for synchrotron polarization). Although the interpretations differed, the results imply that the \planck{} constraint on the $E/B$ power asymmetry suggests a dominance of either isotropic fast waves or sufficiently anisotropic slow and Alfv\'en waves in a low-$\beta$ plasma. \citet{2017ApJ...839...91C} raised the questions whether such conditions are realistic, but \citet{2017MNRAS.472L..10K} argued that the \citet{1995ApJ...438..763G} theory \edit1{and subsequent numerical studies \citep[e.g.,][]{2002PhRvL..88x5001C,2003MNRAS.345..325C} favor such conditions in sub-Alfv\'enic turbulence (i.e., isotropic fast waves and anisotropic slow and Aflv\'en waves with sensitive dependence on the Alfv\'en Mach number),} which appear to be common in the high-latitude sky probed by PXXX. 

It is potentially interesting to attempt to use the $E/B$ power asymmetry and $TE$  correlation coefficient as a probe of turbulence parameters, e.g., Alfv\'en Mach number and plasma $\beta$. For example, \citet{2017MNRAS.472L..10K} claimed that $\REB$ would be higher as the magnetic fields gets stronger in sub-Alfv\'enic, low-$\beta$ turbulence. However, a clear and sensitive dependence of dust polarization statistics to \emph{global} turbulence parameters are not evident in our synthetic maps. Turbulence in our simulations is driven by realistic driving mechanisms (supernovae) and rates (self-regulated by self-consistent feedback loops). The spatially and temporally localized driving interacts with large-scale structures of the ISM developed by gravitational collapse, vertical stratification, thermal instability, and galactic differential rotation. This is drastically different from ``turbulence-in-a-box'' simulations with prescribed driving schemes and isothermal equation of state \citep[e.g.,][]{2002PhRvL..88x5001C,2003MNRAS.345..325C,2010ApJ...720..742K,2010A&A...512A..81F}. Therefore, the characterization of \emph{local} turbulence parameters in our simulation is itself another challenge. A more careful analysis is an essential prerequisite to successfully connect dust polarization statistics with \emph{local} and/or \emph{global} turbulence properties in the multiphase, stratified ISM.

\item  MHD simulations are a promising tool to generate realistic maps for galactic foregrounds, especially for polarized dust emission. Further improvements of foreground maps based on MHD simulations are possible and are currently underway. We are, for example, incorporating the adaptive ray tracing module available in the Athena code \citep{2017ApJ...851...93K} to calculate interstellar radiation fields at every grid point. This will be combined with more sophisticated dust models (Hensley and Draine, in preparation) to obtain a spatially-varying spectral energy distribution and to produce multi-frequency maps. These simulations will allow to assess the degree of frequency decorrelation arising from variation of dust properties in inhomogeneous structures along the line-of-sight, which is a key parameter for future CMB experiments.
\end{enumerate}

\acknowledgements

We are grateful to the referee for helpful report.
CGK acknowledges Eve Ostriker for her continuous encouragement and Juan Soler, Susan Clark, Brandon Hensley, Colin Hill, and Fran\c{c}ois Boulanger for useful discussions. Resources supporting this work were provided in part by the NASA High-End Computing (HEC) Program through the NASA Advanced Supercomputing (NAS) Division at Ames Research Center and in part by the Princeton Institute for Computational Science and Engineering (PICSciE) and the Office of Information Technology's High Performance Computing Center. This work took part under the program Milky-Way-Gaia of the PSI2 project funded by the IDEX Paris-Saclay, ANR-11-IDEX-0003-02. The Flatiron Institute is supported by the Simons Foundation.
CGK and RF are supported in part by the NASA TCAN grant No. NNH17ZDA001N-TCAN.
CGK also acknowledges support from the NASA ATP grant No. NNX17AG26G and from the Simons Foundation Award No. 528307.
RF is grateful for support from the Alfred P. Sloan Foundation, the Department of Energy under grant de-sc0009919, and the NASA ATP Grant No. 80NSSC18K0561. SKC acknowledges support from the Cornell Presidential Postdoctoral Fellowship.

\software{
{\tt Athena} \citep{2008ApJS..178..137S,2009NewA...14..139S},
{\tt yt} \citep{2011ApJS..192....9T}, 
{\tt astropy} \citep{2013A&A...558A..33A}, 
{\tt matplotlib} \citep{Hunter:2007}, 
{\tt numpy} \citep{vanderWalt2011}, 
{\tt IPython} \citep{Perez2007}, 
{\tt pandas} \citep{mckinney-proc-scipy-2010},  
{\tt magnetar} \citep{2013ApJ...774..128S}, 
{\tt healpy} \citep{2005ApJ...622..759G}.}

\bibliographystyle{aasjournal} 
\bibliography{references}{}

\end{document}